\documentclass[12pt,amsmath,amssymb,nofootinbib]{revtex4-1}
\usepackage{amsmath} %
\usepackage{pstricks}
\usepackage{color} 
\usepackage{amssymb}
\input{epsf} 
\def\beq{\begin{equation}} \def\eeq{\end{equation}}
\def\beqn{\begin{eqnarray}} \def\eeqn{\end{eqnarray}}
 \def\to{\rightarrow}

\begin{document} \newcommand\sss{\scriptscriptstyle}
\newcommand\mug{\mu_\gamma} \newcommand\mue{\mu_e}
\newcommand\mui{\mu_{\sss I}} \newcommand\muf{\mu_{\sss F}}
\newcommand\mur{\mu_{\sss R}} \newcommand\muo{\mu_0} \newcommand\me{m_e}
\newcommand\as{\alpha_{\sss S}} \newcommand\ep{\epsilon}
\newcommand\Th{\theta} \newcommand\epb{\overline{\epsilon}}
\newcommand\aem{\alpha_{\rm em}} \newcommand\refq[1]{$^{[#1]}$}
\newcommand\avr[1]{\left\langle #1 \right\rangle}
\newcommand\lambdamsb{\Lambda_5^{\rm \sss \overline{MS}}}
\newcommand\qqb{{q\overline{q}}} \newcommand\qb{\overline{q}} %
\newcommand\MSB{{\rm \overline{MS}}} \newcommand\MS{{\rm MS}}
\newcommand\DIG{{\rm DIS}_\gamma} \newcommand\CA{C_{\sss A}}
\newcommand\DA{D_{\sss A}} \newcommand\CF{C_{\sss F}}
\newcommand\TF{T_{\sss F}} % \newcommand\Jetlist{\{J_l\}_{1,2}}
\newcommand\aoat{\sss a_{\sss 1} a_{\sss 2} a_{\sss 3}}
\newcommand\aoathh{\sss a_{\sss 1} a_{\sss 2} a_{\sss 3} a_{\sss 4}}
\newcommand\SFfull{\{k_l\}_{1,5}}
\newcommand\SFfullbi{\{k_l\}_{1,5}^{[i]}}
\newcommand\SFTfull{\{k_l\}_{1,4}} \newcommand\SFj{\{k_l\}_{3,5}}
\newcommand\SFjh{\{k_l\}_{4,5}} \newcommand\SFjhh{\{k_l\}_{5}}
\newcommand\FLfull{\{a_l\}_{1,5}}
\newcommand\FLfullbi{\{a_l\}_{1,5}^{[i]}}
\newcommand\FLTfull{\{a_l\}_{1,4}} \newcommand\FLFj{\{a_l\}_{3,5}}
\newcommand\FLFjh{\{a_l\}_{4,5}} \newcommand\FLFjhh{\{a_l\}_{5}}
\newcommand\SFjbi{\{k_l\}_{4,5}^{[i]}}
\newcommand\FLjbi{\{a_l\}_{4,5}^{[i]}}
\newcommand\FLjbih{\{a_l\}_{4,5}^{[i]}}
\newcommand\SFjexcl{\{k_l\}_{i,j}^{[np..]}}
\newcommand\SCollfull{\{k_l\}_{1,6}^{[ij]}}
\newcommand\SColl{\{k_l\}_{3,6}^{[ij]}}
\newcommand\FLCollfull{\{a_l\}_{1,6}^{[ij]}}
\newcommand\FLColl{\{a_l\}_{3,6}^{[ij]}} \newcommand\STj{\{k_l\}_{3,4}}
\newcommand\STjh{\{k_l\}_{4}} \newcommand\FLTj{\{a_l\}_{4}}
\newcommand\Argfull{\FLfull;\SFfull}
\newcommand\ArgTfull{\FLTfull;\SFTfull} \newcommand\KtoKF{(k_1,k_2\to
k3,\SFjh\,;\FLFj)} \newcommand\KtoKT{(k_1,k_2\to k3,\STjh\,;\FLTj)}
\newcommand\FLsum{\sum_{\{a_l\}}} \newcommand\FLFsum{\sum_{\FLFjh}}
\newcommand\FLFsumhh{\sum_{\FLFjhh}}
\newcommand\FLFsumbi{\sum_{\FLjbih}} \newcommand\FLTsum{\sum_{\FLTj}}
\newcommand\MF{{\cal M}^{(3,0)}} \newcommand\MT{{\cal M}^{(2,0)}}
\newcommand\MTz{{\cal M}^{(2,0)}} \newcommand\MTo{{\cal M}^{(2,1)}}
\newcommand\MTi{{\cal M}^{(2,i)}} \newcommand\MTmn{{\cal
M}^{(2,0)}_{mn}} \newcommand\MFsj{\MF(k_1,k_2\to k3,\SFjh)}
\newcommand\MTsj{\MT(k_1,k_2\to k3,\STjh)}
\newcommand\MTisj{\MTi(k_1,k_2\to k3,\STjh)}
\newcommand\PHIFsj{\phi_3(k_1,k_2\to k_3,\SFjh)}
\newcommand\PHIFsjhh{\phi_3(k_1,k_2\to k_3,k_4,\SFjhh)}
\newcommand\PHITsj{\phi_2(k_1,k_2\to k_3,\STjh)}
\newcommand\PHITsjhh{\phi_2(k_1,k_2\to k_3,k_4)}
\newcommand\uoffct{\frac{1}{2!}} \newcommand\uotfct{\frac{1}{1!}}
\newcommand\uoxic{\left(\frac{1}{\xi}\right)_c}
\newcommand\uoxiic{\left(\frac{1}{\xi_i}\right)_c}
\newcommand\uoxilc{\left(\frac{\log\xi}{\xi}\right)_c}
\newcommand\uoxiilc{\left(\frac{\log\xi_i}{\xi_i}\right)_c}
\newcommand\uoyim{\left(\frac{1}{1-y_i}\right)_+}
\newcommand\uoyimdi{\left(\frac{1}{1-y_i}\right)_{\delta_{\sss I}}}
\newcommand\uoyjmdo{\left(\frac{1}{1-y_j}\right)_{\delta_o}}
\newcommand\uoyip{\left(\frac{1}{1+y_i}\right)_+}
\newcommand\uoyipdi{\left(\frac{1}{1+y_i}\right)_{\delta_{\sss I}}}
\newcommand\uoyilm{\left(\frac{\log(1-y_i)}{1-y_i}\right)_+}
\newcommand\uoyilp{\left(\frac{\log(1+y_i)}{1+y_i}\right)_+}
\newcommand\uozm{\left(\frac{1}{1-z}\right)_+}
\newcommand\uozlm{\left(\frac{\log(1-z)}{1-z}\right)_+}
\newcommand\SVfact{\frac{(4\pi)^\ep}{\Gamma(1-\ep)}
\left(\frac{\mu^2}{Q^2}\right)^\ep} % \newcommand\gs{g_{\sss S}}
\newcommand\Icol{\{d_l\}} \newcommand\An{{\cal A}^{(n)}}
\newcommand\Mn{{\cal M}^{(n)}} \newcommand\Nn{{\cal N}^{(n)}}
\newcommand\Anu{{\cal A}^{(n-1)}} \newcommand\Mnu{{\cal M}^{(n-1)}}
\newcommand\Sumae{\sum_{d_{e}}} \newcommand\Sumaecl{\sum_{d_{e},\Icol}}
\newcommand\Sumaeae{\sum_{d_{e},d_{e}^{\prime}}}
\newcommand\Sumhe{\sum_{h_{e}}} \newcommand\Sumhep{\sum_{h_{e}^\prime}}
\newcommand\Sumhehe{\sum_{h_{e},h_{e}^{\prime}}}
\newcommand\Pgghhh{S_{gg}^{h_e h_i h_j}} \newcommand\Pggplus{S_{gg}^{+
h_i h_j}} \newcommand\Pggminus{S_{gg}^{- h_i h_j}}
\newcommand\Pgghphh{S_{gg}^{h_e^\prime h_i h_j}}
\newcommand\Pqgplus{S_{qg}^{+ h_i h_j}} \newcommand\Pqgminus{S_{qg}^{-
h_i h_j}} \newcommand\Pgqplus{S_{gq}^{+ h_i h_j}}
\newcommand\Pgqminus{S_{gq}^{- h_i h_j}} \newcommand\Pqqplus{S_{qq}^{+
h_i h_j}} \newcommand\Pqqminus{S_{qq}^{- h_i h_j}}
\newcommand\Hnd{\{h_l\}} \newcommand\LC{\stackrel{\sss i\parallel
j}{\longrightarrow}} \newcommand\LCu{\stackrel{\sss 1\parallel
j}{\longrightarrow}} % \newcommand\Physvar{\{V_l\}}
\newcommand\Physcm{\{\bar{V}_l\}} \newcommand\Partvar{\{v_l\}}
\newcommand\Partcm{\{\bar{v}_l\}}
\newcommand\Partvarcm{\{v_l(\bar{v})\}}
\newcommand\Partcmset{\{\bar{v}_l^{(i)}\}}
\newcommand\Partvarcmset{\{v_l(\bar{v}^{(i)})\}}

\begin{titlepage} 
\renewcommand{\thefootnote}{\fnsymbol{footnote}}
\begin{flushright} 
% \\ hep-ph/yymmnnn 
\end{flushright} %\par
\vspace{10mm}

\begin{center} 
\Large{ {\bf Hadron plus photon production in polarized hadronic\\} 
\vspace*{0.3cm} {\bf collisions at next-to-leading order
accuracy\\}} 
\vskip 1.cm {\large  Daniel de Florian and Germ\'an F.R.
Sborlini }  \\
\vspace*{0.3cm} 
\normalsize Departamento de F\'\i sica\\
Facultad de Ciencias Exactas y Naturales \\ Universidad de Buenos Aires
\\ Pabell\'on I, Ciudad Universitaria \\ (1428) Capital Federal \\
Argentina \end{center} 
\vskip 1.cm 
\begin{center}
{\bf Abstract} \\
\end{center}
\vspace{0.2cm} \normalsize

We compute the next-to-leading order QCD corrections to the polarized (and unpolarized) cross sections for the production of a hadron accompanied by an opposite-side prompt photon. This process, being studied at RHIC, permits us to reconstruct partonic kinematics using experimentally measurable variables. We study the correlation between the reconstructed momentum fractions and the true partonic ones, which in the polarized case might allow us to reveal the spin-dependent gluon distribution with a higher precision.
\\ 
February 2011
\end{titlepage} 
\section{Introduction}

It is well known that only a small fraction of the spin of the proton is carried by quarks. This information was gathered through the study of longitudinally polarized deep-inelastic scattering (DIS) processes. However DIS data is not enough to extract the full $x$-dependence of the polarized quark and gluon densities of the nucleon \cite{deFlorian:2008mr}. Furthermore, the $x$-shape of $\Delta g$ seems to be hardly constrained by DIS data because it contributes in the leading order (LO) only through the $Q^{2}$ dependence of $g_{1}$.

The precise extraction of $\Delta g$ thus remains one of the most interesting challenges for spin physics experiments. Unlike DIS,  processes involving the collision of polarized protons, like at the RHIC collider at BNL, introduce a direct gluonic contribution already at the lowest order in perturbation theory, resulting in a very useful tool to extract more information about $\Delta g$.

A recent next-to-leading-order (NLO) global analysis, performed including all available inclusive and semi-inclusive DIS and RHIC data, suggests that the gluon polarization in the nucleon is rather small in the region $0.05<x<0.2$ \cite{deFlorian:2008mr,deFlorian:2009vb}. Measured observables in hadronic collisions include single-pion production at center-of-mass energy $\sqrt{s}=$ 62 \cite{Adare:2008qb} and 200 GeV  \cite{Adare:2007dg}, and jet production at $\sqrt{s}=$ 200 GeV \cite{Abelev:2007vt}.

Both experimental collaborations at RHIC have started to study less inclusive observables that might turn out to be useful to pin down the shape of the polarized parton distributions with better precision. For example, the production of a charged hadron accompanied by a back-to-back jet or prompt photon allows experimentalists to use the last one as a trigger for hadron detection, reducing the bias in the selection of events \cite{hjet}. Furthermore, from a theoretical point of view, counting with an extra particle (like a direct photon) in the final state provides a cleaner resource to access parton kinematics. The dominance of the partonic $qg\rightarrow \gamma q$ channel in the case of photon production supplies a direct signal on the gluon distribution in polarized collisions.

Looking at a photon as a trigger has a relevance also in heavy ion collisions. When a hot and dense high-energy medium is formed, QCD particles interact strongly with this medium, causing them to lose energy as observed in Refs. \cite{Adare:2009vd,Adare:2010yw,Abelev:2009gu}. Since the photon weakly interacts with that medium, this particle might travel through it without being strongly perturbed. For that reason, observing direct photons in coincidence with hadrons turns out to be useful to reconstruct the parton kinematics in heavy ion collisions as well and, therefore, to establish the pattern of partonic energy loss \cite{Wang:1996yh}. Recent phenomenological analysis \cite{Zhang:2009rn,Arleo:2004xj,Arleo:2006xb} indicate that studying distributions in the fragmentation fraction variable $z$ is particularly sensitive to the medium modified fragmentation functions at RHIC.

Furthermore, a similar observable corresponding to the production of a jet and a photon can be used, as proposed in \cite{Belghobsi:2009hx}, to constrain the photon fragmentation functions at RHIC in a region barely accessible by LEP kinematics.

In order to make reliable quantitative predictions for a high-energy process, it is crucial to determine the NLO QCD corrections to the Born approximation. In hadronic collisions, cross sections computed at the lowest order in perturbation theory are only of qualitative value, since higher order corrections are usually sizable, in some cases as large as or even larger than the Born contribution. Besides that, the LO calculation is severally affected by the dependence on the "`unphysical"' factorization and renormalization scales, dependence that can be partially cured only by including the higher order corrections. Also, for less inclusive processes, the inclusion of extra partons in the NLO perturbative calculation allows one to improve the matching between the theoretical result and the realistic experimental conditions. Furthermore, the usual "`naive"' relations between the momentum of the outgoing particles and the kinematics of the participating partons in the hard process are usually exact only at leading order accuracy, being mandatory to establish their range of validity when higher order corrections are taken into account. NLO corrections for $\gamma + h$ were computed for unpolarized collisions in \cite{Binoth:2002wa}, where the corrections are found to be large, on the order of $70\%$, even for LHC energies.

The purpose of this paper is, therefore, to extend the available calculations by computing the NLO corrections to hadron plus prompt-photon production in polarized hadronic collisions and to analyze its phenomenological impact. Nowadays there are several versions of the subtraction method that allow the
computation of any infrared-safe quantity to NLO accuracy. We rely on the formalism of Ref. \cite{Frixione:1995ms}, which has been used previously to compute observables in jet \cite{deFlorian:1998qp,deFlorian:1999ge}, prompt-photon \cite{frixione-etal}, single-hadron \cite{singlehad} and hadron+jet production\cite{deFlorian:2009fw}, for both polarized and unpolarized cases.

Most of the advantages for detecting a photon discussed above rely on the fact that the photon couples directly (pointlike) to the partons in the hard process and propagates as a colorless state. Unfortunately, this is true only for {\it direct} photons, and the cleanliness of the process is spoiled by {\it resolved} photons that can be produced from a parton in the QCD reaction that fragments in a photon plus a number of hadrons. Several isolation algorithms have been proposed to reduce this contribution. The main idea in all of them is that nondirect photons are surrounded by hadronic debris, so measuring the amount of hadronic energy around a photon allows us to define a selection criteria. In this paper we apply the isolation prescription introduced by Frixione in Ref. \cite{Frixione:1998jh}, which has the theoretical advantage that, still being infrared-safe, completely eliminates the unwanted resolved
component to the cross section. 

Relying on this isolation algorithm, we construct a Monte Carlo code that calculates any infrared-safe observable to NLO precision in hadron plus direct photon production. The code provides results for both unpolarized and polarized
collisions and allows us to impose experimental cuts and to perform a detailed phenomenological analysis. Specifically, we use the Monte Carlo code to explore the sensitivity of some observables on the spin-dependent gluon distributions in
polarized proton-proton collisions. Also, we study correlations between certain sets of external experimentally-accessible variables and real partonic momentum fractions in the unpolarized case.

The paper is organized as follows: in Sec. II  we describe the theoretical tools used to perform the NLO calculation and study the perturbative stability of the LO and NLO results. In Sec. III we analyze some general aspects of unpolarized and polarized collisions. In Sec. IV, we focus on the polarized case, showing predictions for asymmetries at different center-of-mass energies and both RHIC experiments. Finally, we present the conclusions in Sec. V.

\section{NLO corrections and Perturbative stability}

The factorization theorem \cite{CSS} allows one to separate the nonperturbative effects associated with the hadronic structure from those calculable within perturbation theory and, therefore, to write the hadronic cross section as 
\beqn 
d\sigma^{(H_1 H_2 \rightarrow h\gamma)}(K_1,K_2,K_3,K_4)&=&\sum_{a_1 a_2 a_3}\int dx_1 dx_2 dz \,
f^{(H_1)}_{a_1}(x_1,{\mu}_I)\, f^{(H_2)}_{a_2}(x_2,{\mu}_I)
D_{a_3}^{(h)}(z, {\mu}_F) \nonumber \\ &&
d\hat{\sigma}_{\aoat{\gamma}}(x_1 K_1,x_2 K_2, K_3/z, K_4; \mur, \mui,
\muf)\ \nonumber \\
+ \sum_{a_1 a_2 a_3 a_4}\int dx_1 dx_2 \nonumber  dz_1
dz_2 && \hspace{-0.4cm} f^{(H_1)}_{a_1}(x_1, {\mu}_I) \, f^{(H_2)}_{a_2}(x_2, {\mu}_I)\,
D_{a_3}^{(h)}(z_1, {\mu}_F)\, D_{a_4}^{(\gamma)}(z_2, {\mu}_F)
\nonumber \\ && d\hat{\sigma}_{\aoathh}(x_1 K_1,x_2 K_2, K_3/z_1,
K_4/z_2; \mur, \mui, \muf) \, , \label{factthNLOFO} 
\eeqn 
where $H_1$ and $H_2$ are the colliding hadrons, $K_1$ and $K_2$ their four-momentum, and the sum is performed over the parton flavors which can contribute to the process. Here, $K_3$ is the final state hadron four-momentum,
while $K_4$ denotes the photon momentum. The first term represents the {\it direct} contribution whereas the second one accounts for the {\it resolved} component \footnote{Notice that in general the decomposition between direct and resolved contributions from NLO accuracy is not physical. Nevertheless, with the isolation prescription used along this work only the first term survives, which implies that the direct contribution to the total cross section can be unambiguously defined.}. $d\hat{\sigma}_{\aoat \gamma}$ and
$d\hat{\sigma}_{\aoathh}$ are the corresponding subtracted partonic cross sections, in which collinear singularities have been removed through the introduction of suitable counterterms. $f^{(H_i)}_{a_j}$ is the partonic distribution function (PDF) for a parton $a_j$ inside an hadron $H_i$, while $D_{a_i}^{(h)}$ denotes the fragmentation function associated with a parton $a_i$ which hadronises into a hadron $h$. In the {\it resolved} term   also the necessary $D_{a_i}^{(\gamma)}$ {\it photon fragmentation function} appears. The expression in Eq. (\ref{factthNLOFO}) implicitly assumes that the photon and the final state hadron are well separated in angle, to avoid extra collinear configurations that would require the introduction of double fragmentation functions \cite{doublefrag}. Here, ${\mu}_I$ and ${\mu}_F$ correspond to the  initial and final state factorization scales, while ${\mu}_R$ is the renormalization scale. The full set of NLO QCD corrections to both direct and resolved contributions were first computed for unpolarized collisions in Ref. \cite{Binoth:2002wa}. The analogous of (\ref{factthNLOFO}) for polarized cross section is obtained by replacing the parton distributions and the partonic cross section by its polarized expressions, $\Delta f^{(H_i)}_{a_j}$ and $d \Delta\hat{\sigma}$, respectively. As usual, the longitudinally polarized asymmetry is defined by the ratio between the polarized and unpolarized
cross sections 
\begin{eqnarray}
{A}_{LL}^{h}=\frac{d\Delta\sigma}{d\sigma} \, . \label{asypt}
\end{eqnarray}

As stated in the introduction, resolved contributions take into account those events in which the photon is generated from parton fragmentation. In order to eliminate its contribution, we rely on the photon isolation prescription introduced in Ref. \cite{Frixione:1998jh}. This algorithm can be summarized in four steps: 
\begin{itemize} 
\item Identify the photon in the final state and define a cone surrounding it with a radius $r_{0}$ in the rapidity-azimuthal plane; 
\item If all QCD partons lie outside the cone, then the photon is isolated; 
\item If there are QCD partons inside the cone, calculate their distance to the
photon 
\beq 
r_{i} = \sqrt{{(\eta_i - \eta_{\gamma})}^2 + {(\phi_i - \phi_{\gamma})}^2} , \eeq 
and define
\beq 
E_T(r) = \sum_{i} E_{T_i} \, \theta{\left(r-r_i\right)} \, ,
\eeq 
which is the total transverse  hadronic energy accumulated in a cone of radius $r$ around the photon; 
\item Introduce an arbitrary function $\xi(r)$ satisfying $\lim_{r\to 0} \xi(r) = 0$. Then the photon will be isolated if $E_T(r) \leq \xi(r)$, for every $r\leq r_0$. 
\end{itemize} 
Notice that this prescription does not forbid the emission of soft gluons in any region of the phase space, but it does eliminate the divergences arising from the kinematical configuration where a quark becomes collinear to the photon. For this reason, the prescription does not spoil the cancellation of infrared singularities and can be applied to any infrared-safe observable.

After the implementation of the isolation criteria only the direct contribution to the cross section survives as 
\beqn 
d\sigma^{(H_1 H_2\rightarrow h \gamma)(ISO)}(K_1,K_2,K_3,K_4)&=&\sum_{\aoat}\int dx_1
dx_2 dz \, f^{(H_1)}_{a_1}(x_1,\mui) f^{(H_2)}_{a_2}(x_2,\mui)
D_{a_3}^{(h)}(z, \muf) \nonumber \\*&& d\hat{\sigma}^{(ISO)}_{\aoat
\gamma}(x_1 K_1,x_2 K_2, K_3 / z, K_4; \mur, \mui, \muf)\ ,
\label{ContribucionISO} 
\eeqn
where the $(ISO)$ superscript indicates that the isolation algorithm has been included in the corresponding measurement function. Following the procedure detailed in Ref. \cite{Frixione:1995ms}, the calculation is implemented in a
Monte Carlo-like code that allows us to compute any infrared-safe observable for the process $H_1 H_2 \rightarrow h \gamma$. It is worth noticing that the same code computes both unpolarized and polarized cross sections, since the formal structure of the corrections is exactly the same.

In this work, we mostly concentrate on the phenomenology of charged pion production accompanied by a back-to-back photon. Our default results will correspond to the kinematics of the PHENIX experiment at RHIC with a center-of-mass energy of $\sqrt{s}=200$ GeV, but we will also present asymmetries at $\sqrt{s}=500$ GeV and for the kinematics corresponding to the STAR experiment as well. Unless otherwise stated, we require the pion transverse momenta to be larger than $2$ GeV and the one for the photon to be restricted to the $5\, {\rm GeV} < p_T^{\gamma} < 15$ GeV range. These cuts are imposed to take into account experimental limitations to identify direct photons \cite{Adare:2009vd,Adare:2010yw}.

The rapidities of the pion and the photon are limited to the range $|\eta|<0.35$, according to the constrains imposed by the PHENIX detector \footnote{For the sake of simplicity we assume full azimuthal coverage for our phenomenological results. The rapidity limits are set to $|\eta|<1$ for STAR.}. The particles have to be separated in the azimuthal angle by $\Delta \phi\equiv |\phi^{\pi}-\phi^{\gamma}| > 2$ to ensure the pion is produced from a parton in the `opposite side' to the photon. For the isolation algorithm we chose the radius as $r_0 = 0.4$, the functional form 
\beqn 
\xi(r) &=& \epsilon_{\gamma}  E_T^{\gamma}
\, \left(\frac{1-\cos(r)}{1-\cos(r_0)} \right)^n \, ,
\label{ChiEXPRESSION} 
\eeqn
and set the parameters in Eq. (\ref{ChiEXPRESSION}) by $n=4$ and $\epsilon_\gamma = 1$.

Since there are two different hard scales in the process, given by the transverse momenta of the pion and the photon, the default initial $(\mu_I)$ and final $(\mu_F)$ state  factorization scales and the renormalization scale $(\mu_R)$ are chosen as their average. Therefore, unless otherwise stated the scales are defined as 
\beq 
\mu_I=\mu_F=\mu_R = \mu\equiv\frac{p^{\gamma}_T + p^{\pi}_T}{2} \, .
\eeq 
In the unpolarized case, we rely on the MRST2002 parton distribution functions (PDFs) \cite{mrst2002}, since it is the reference set for the DSSV polarized PDFs that we will use as the default density in the computation of the corresponding asymmetry. For the fragmentation functions we rely on the DSS \cite{deFlorian:2007hc} set that provides full flavor and charge separation at NLO. We will later discuss the consequences of using different sets of parton distributions and fragmentation functions.

%%==================================== 
\begin{figure}[htb]
\vspace{-0.3cm} 
\begin{center} 
\begin{tabular}{c} 
\epsfxsize=14truecm \epsffile{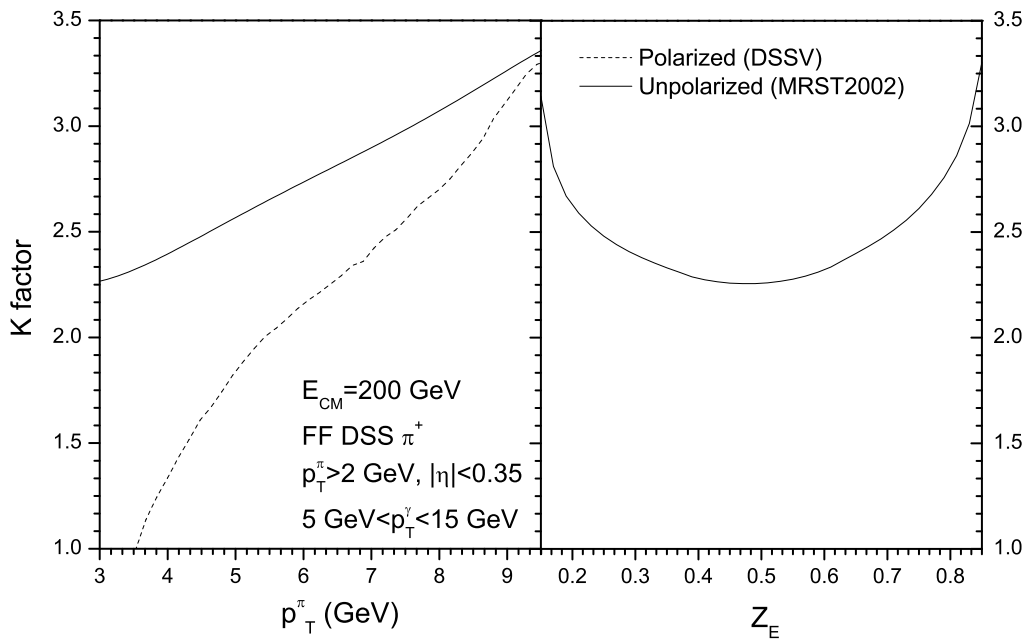}\\ 
\end{tabular} 
\end{center}
\vspace{-0.5cm} \caption{{\em \label{fig:K}  Unpolarized (solid line) and polarized (dashed line) NLO K-factors for the pion transverse momentum distribution (left) and the variable $Z_E$ (right). The choice of the factorization and renormalization scales corresponds to the default value
$\mu_I=\mu_F=\mu_R=\mu\equiv\left( p_T^{\pi} +p_T^{\gamma}\right)/2$. The polarized K-factor is not presented in the right-side plot since the polarized cross section has nodes in that range of $Z_E$.}}
\end{figure}
%%==================================== 

It is important to remark that we will always use NLO PDFs and fragmentation functions for both LO and NLO calculations. This choice is particularly important in the polarized case where the uncertainty in the distributions can render artificially large corrections because of the differences between the LO
and NLO polarized densities. In other words, any difference we find between LO and NLO cross sections will arise only from the corresponding partonic cross section.  Also, we use in both cases the two-loop expression for $\as$, and the one-loop correction to $\alpha$.

A useful way to quantify the size of the QCD corrections is by means of 
the $K$-factor, defined as the ratio between the NLO and LO results. In
Fig. \ref{fig:K} we show the $K$-factor as a function of $p_{T}^{\pi}$ and of the scaling variable $Z_E$ \footnote{This variable is denoted as $X_E$ in Refs. \cite{Adare:2009vd,Adare:2010yw}.} defined as 
\beq
Z_{E} = -\cos{\left(\phi^{\pi}-\phi^{\gamma}\right)} 
\frac{p^{\pi}_T}{p^{\gamma}_T} \, ,
\label{DefinicionZE}
\eeq
where $\phi^{\pi}$ and $\phi^{\gamma}$ are the azimuthal angle associated to the pion and the photon, respectively. We will discuss in the next section the
usefulness of this variable, which at lowest order represents the argument ($z$) of the fragmentation function in Eq. (\ref{ContribucionISO}).
%%==================================== 
\begin{figure}[htb]
\vspace{-0.3cm} 
\begin{center} 
\begin{tabular}{c} 
\epsfxsize=14truecm \epsffile{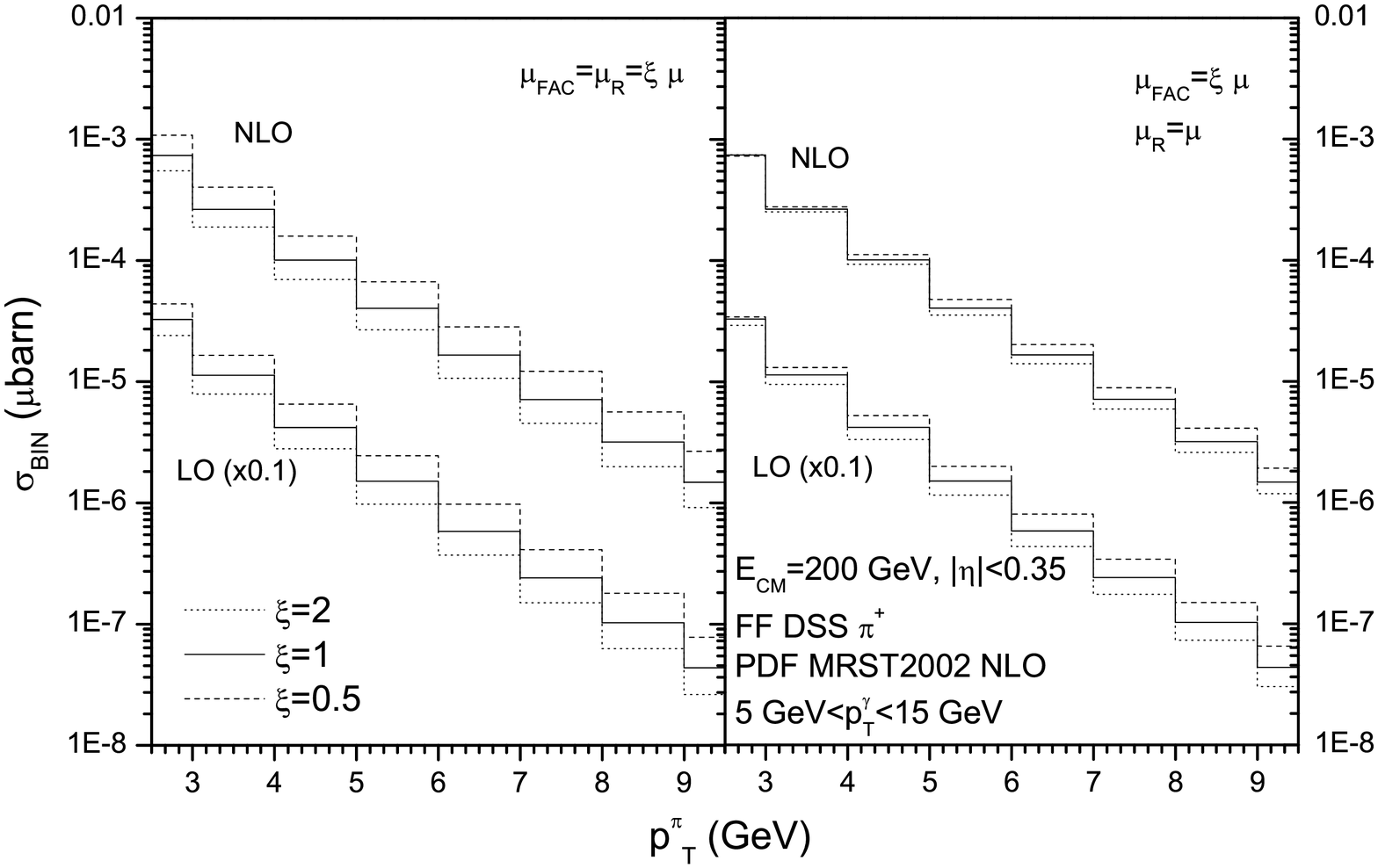}\\ 
\end{tabular} 
\end{center}
\vspace{-0.5cm} \caption{{\em \label{fig:scales1} Scale dependence of the  unpolarized cross section in terms of the transverse momentum of the pion. The left side corresponds to the simultaneous
variation of factorization and renormalization scales as $\mu_I=\mu_F=\mu_R= \xi \mu \equiv \xi \left( p_T^{\pi} +p_T^{\gamma}\right)/2$, while the right side shows the result when the renormalization scale is kept fixed to the default value. }}
\end{figure}
%%====================================
As it happens in the case of single-hadron production \cite{singlehad,Jager:2002xm}, the NLO corrections turn out to be very large, with a $K$-factor that lies in the range 2.2 $\leq K \leq$ 3.2, and with more moderate (but still sizable) corrections for the polarized cross section at lower transverse momentum. Notice that for these observables the $K$-factor is a nontrivial function of $p_T^{\pi}$ and $Z_E$, respectively. Furthermore, the pattern for the QCD corrections in the polarized case considerably differs from the unpolarized one, which will result in non-negligible corrections for the asymmetries.

Besides the $K$-factor study, we need to estimate the corresponding theoretical uncertainty for this calculation. Because we are not able to evaluate the full-perturbative expansion, we can indirectly test the quality of the NLO expansion through a scale-dependence analysis. As it is well known, factorization and renormalization scales are not physical parameters of the real problem, so the exact result must not depend on them. But theoretical predictions do have such a dependence, arising from the truncation of the perturbative expansion at a fixed order in the coupling constant $\alpha_s$. A large dependence on the scales, therefore, implies a large theoretical uncertainty. A full NLO analysis of scale dependence for unpolarized collisions was presented in Ref. \cite{Binoth:2002wa}. Nevertheless, and for the sake of completeness, before showing the new results for the polarized cross section we recall the main features in the unpolarized case at RHIC kinematics.
%%==================================== 
\begin{figure}[h]
\vspace{-0.3cm} 
\begin{center} 
\begin{tabular}{c} 
\epsfxsize=14truecm \epsffile{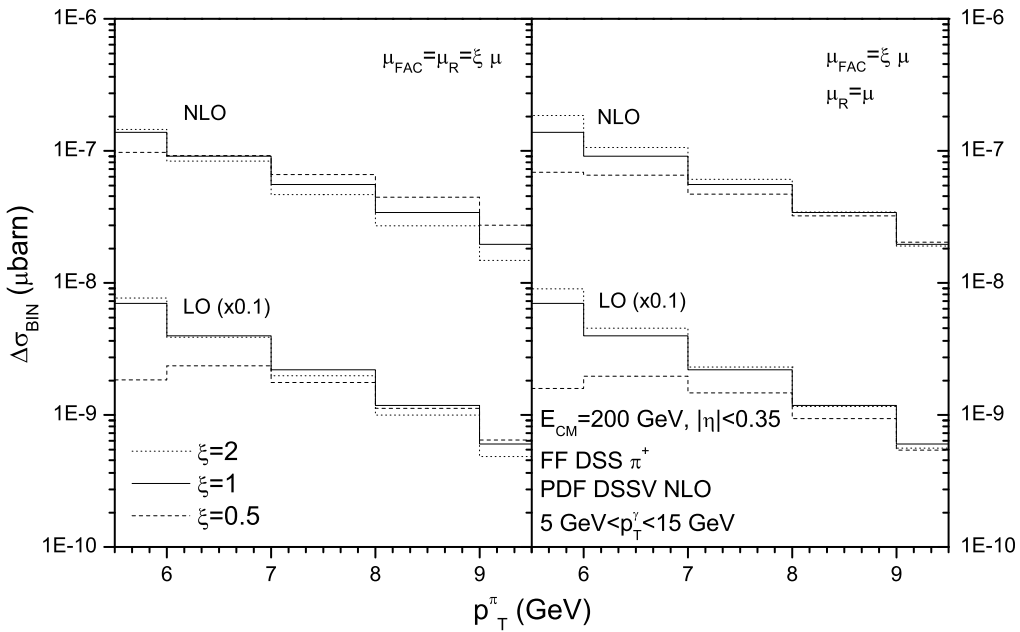}\\ 
\end{tabular} 
\end{center}
\vspace{-0.5cm} \caption{{\em \label{fig:scales2} Same as Fig. \ref{fig:scales1}
but for the polarized cross section. The plot starts in this case at $p^{\pi}_T=6 \, {\rm GeV}$ since at lower values the polarized cross section changes sign.}}
\end{figure}
%%====================================
As a first approach to analyze the perturbative stability of the observable through the scale dependence, we set again all factorization and renormalization scales to be equal and vary them by a factor of 2 up and down with respect to the default choice, i.e., $\mu_I=\mu_F=\mu_R=\xi \left( p_T^{\pi} +p_T^{\gamma} \right)/2$ with $\xi=1/2,1,2$. The left side of Fig. \ref{fig:scales1} plots the corresponding scale dependence of the LO and NLO unpolarized cross sections. When the three scales are varied together, both LO and NLO results exhibit a very similar ${\cal O}(\pm 50\%)$ variation, which is at first sight  disappointing since one expects an improvement after including the NLO corrections. However, this is not an indication of a failure of the perturbative expansion as it might seem. On one hand, a large scale dependence is not truly unexpected since similar results are found for single-hadron
production \cite{singlehad,Jager:2002xm}. The fact that no visible improvement is found at NLO for this choice of scales has also been observed in the case of inclusive prompt-photon production \cite{frixione-etal}, which relies on the same matrix elements used for this calculation. This trend can be attributed to the particular choice of scales, produced because of an incidental cancellation between the effects of the factorization and renormalization scale variations at LO when they are varied simultaneously.

Therefore, a better way to analyze the issue of perturbative stability in this case is to perform an independent variation of the scales. As an example, we keep the renormalization scale fixed to the default value (i.e., $\mur=\mu$) and vary only the initial and final state factorization scales in the way described above ($\mu_I=\mu_F=\xi \mu$). The result of this study is shown in the right side of Fig. \ref{fig:scales1}, where we do observe that the LO cross
section presents an average fluctuation of $\pm$ 30$\%$ whereas the corresponding NLO result varies only about $\pm$ 15$\%$. Thus, in this scenario we can assert that NLO is truly improving the accuracy of the calculation. The dependence on the renormalization scale, when the factorization scales are fixed to the default value, remains almost the same at LO and NLO. The lack of reduction on the $\mu_R$ dependence at NLO is due to the isolation cuts which, while being an infrared-safe prescription, still perturb the cancellation of the soft-gluon effects and therefore have an impact on the renormalization scale dependence \cite{frixioneIchep}. Most of the scale dependence, particularly the one arising from the renormalization scale, will cancel in the cross section ratios built in order to analyze the nuclear effects, and in the asymmetries defined in polarized collisions as well. Nevertheless, despite the improvements at NLO that confirm the perturbative stability, it is clear that the theoretical uncertainty remains rather sizable for the cross sections themselves \footnote{Variations of the order of $40\%$ are quoted in Ref. \cite{Binoth:2002wa} for the $\pi\gamma$ invariant mass distributions at the LHC after implementing a different isolation prescription.}. All the results in Fig. \ref{fig:scales1} correspond to the pion transverse momentum distributions. We observe similar features when we study the scale dependence
of the unpolarized cross section as a function of the {\it scaling} variable $Z_{E}$. The trend of the scale dependence turns out to be the same when the variations are performed around a different default choice, like the invariant mass of the photon-pion pair or different combinations of their transverse momentum like $p_T^{\gamma}$, $p_T^{\pi}$ and ${\rm max} \left( p_T^{\gamma}, p_T^{\pi}\right)$. In particular, we do not find any reasonable choice that presents a visible  decreases in the scale dependence.

%%==================================== 
\begin{figure}[htb]
\vspace{-0.3cm} 
\begin{center} 
\begin{tabular}{c} 
\epsfxsize=14truecm \epsffile{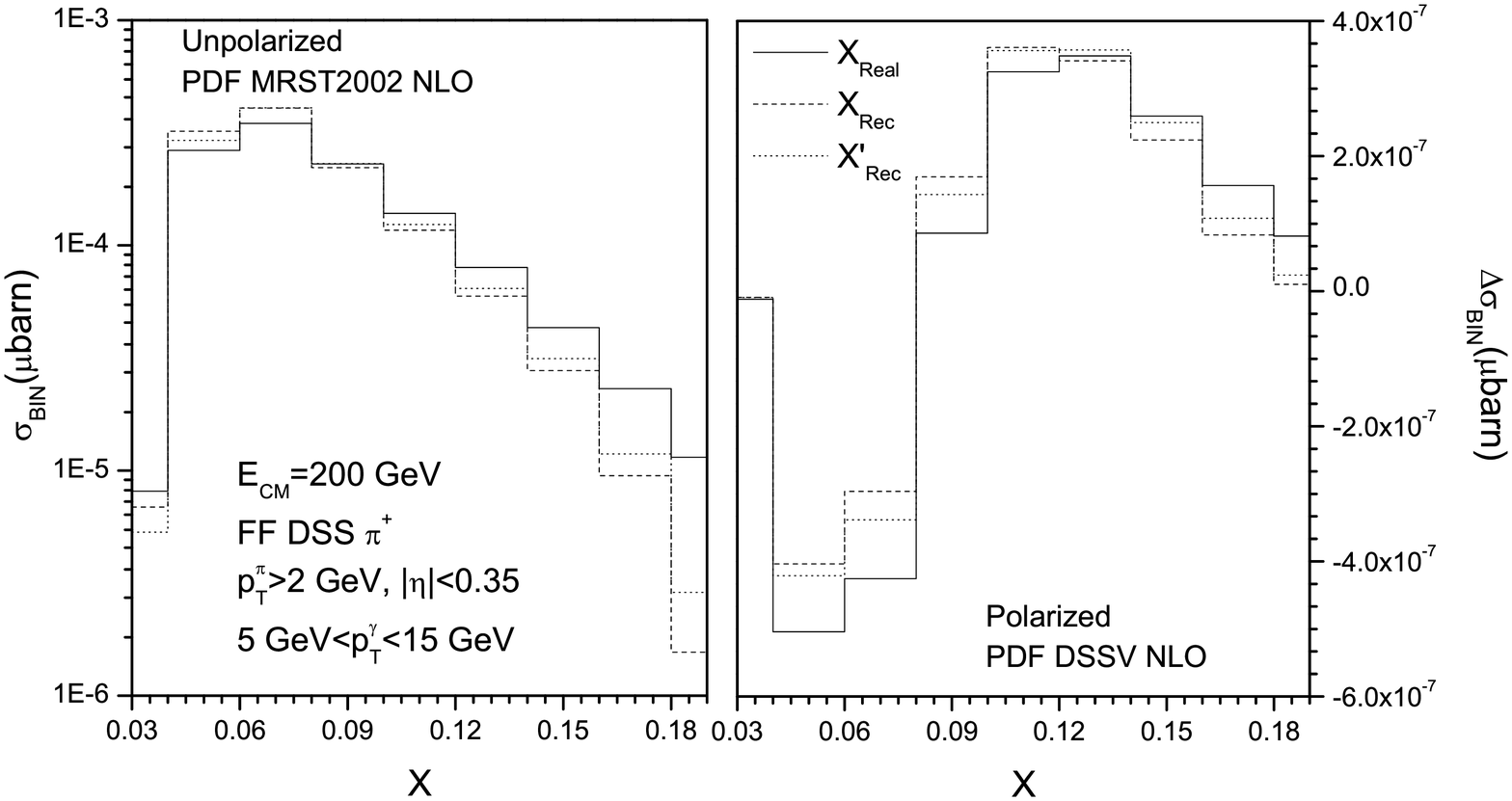}\\ 
\end{tabular} 
\end{center}
\vspace{-0.5cm} \caption{{\em \label{fig:sigmax} 
NLO cross section as a function of $X_{\rm{Real}}$ (solid lines),
$X_{\rm{Rec}}$ (dashed lines) and $X'_{\rm{Rec}}$ (dotted lines), for the polarized (right) and unpolarized (left) cases. Factorization and renormalization scales are set to the default value.}}
\end{figure}
%%====================================

For polarized collisions, the reduction in the scale dependence at NLO is much more evident, as shown in Fig. \ref{fig:scales2}. This feature has been observed for many observables in hadronic collisions and can be mainly attributed to the less singular behavior of polarized parton distributions and the corresponding evolution kernels that drive the initial state factorization scale-dependence.

%%%%%%%%%%%%%%%%%%%%%%%%%%%%%%%%%%%%%%%%%%%%%%%%%%%%%%%%%%%%%%%%
\section{Phenomenology}
%%%%%%%%%%%%%%%%%%%%%%%%%%%%%%%%%%%%%%%%%%%%%%%%%%%%%%%%%%%%%%%%

As discussed in the introduction, counting with the photon and pion kinematics allows us to attempt for a reconstruction, to some degree of accuracy, of the partonic momentum fractions $x_{1,2}$ and $z$. It is worth noticing that at the Born level, the momentum fractions $x_{1,2}$ can be written as 
\beqn
X_{1,\rm{Rec}} &=& \frac{p_{T}^{\gamma}\exp{\left(\eta^{\pi}\right)}+p_{T}^{\gamma}
\exp{\left(\eta^{\gamma}\right)}}{\sqrt{s}} \\X_{2,\rm{Rec}} &=&
\frac{p_{T}^{\gamma}\exp{\left(-\eta^{\pi}\right)}+p_{T}^{\gamma}
\exp{\left(-\eta^{\gamma}\right)}}{\sqrt{s}} \, ,
\eeqn
or equivalently at this level, as defined by the experimental collaborations,
\beqn
 X'_{1,\rm{Rec}} &=&
\frac{p_{T}^{\gamma}\exp{\left(\eta^{\pi}\right)}-\cos{\left(\phi^{\pi}-
\phi^{\gamma}\right)}p_{T}^{\gamma}\exp{\left(\eta^{\gamma}\right)}}
{\sqrt{s}}
\\ X'_{2,\rm{Rec}} &=&
\frac{p_{T}^{\gamma}\exp{\left(-\eta^{\pi}\right)}-\cos{\left(\phi^{\pi}
-\phi^{\gamma}\right)}p_{T}^{\gamma}\exp{\left(-\eta^{\gamma}\right)}}{\sqrt{s}} \, .
\eeqn
At the Born level, these variables are equal to the initial partons momentum fractions, $x_1$ and $x_2$ respectively, because only the $2\to2$ partonic process enter in the calculation and $\cos{\left(\phi^{\pi}-\phi^{\gamma}\right)}=-1$ for the trivial back-to-back configuration. However, NLO computations require the inclusion of virtual and real corrections, the last being associated with $2\to3$ partonic processes. Therefore $X_{i,\rm{Rec}}$ and $X'_{i,\rm{Rec}}$ may differ from the real momentum fraction $x_i$ and between themselves. Counting with the Monte Carlo code presented in the previous sections, we can analyze to what extent these naive relations are modified by the NLO corrections. Recalling the invariance of the physical problem under the exchange of partons 1 and 2 at RHIC, we study the cross section distribution in terms of  the reconstructed variables $X_{\rm{Rec}} \equiv X_{1,\rm{Rec}}$, $X'_{\rm{Rec}} \equiv X'_{1,\rm{Rec}}$ and also the true momentum fraction $X_{\rm{Real}}\equiv x_1$.

%%==================================== 
\begin{figure}[htb]
\vspace{0.5cm} 
\begin{center} 
\begin{tabular}{c} 
\epsfxsize=14truecm \epsffile{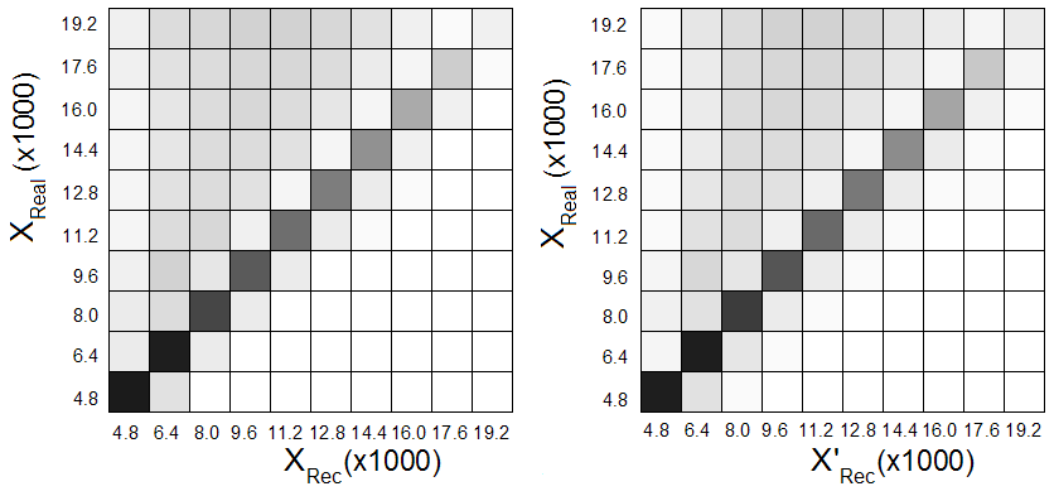}\\ 
\end{tabular} 
\end{center}
\vspace{-0.5cm} \caption{{\em \label{fig:corx} 
Correlation for $X_{\rm{Rec}}$ (left) and $X'_{\rm{Rec}}$
(right) versus $X_{\rm{Real}}$. We use a gray scale picture to indicate
the probability of finding an event with a certain value of
$X_{\rm{Real}}$ and $X_{\rm{Rec}}$ (or $X'_{\rm{Rec}}$): the darker the
bin, the larger the correlation between the variables. For the sake of presentation we plot values of $X$ multiplied by $1000$. }}
\end{figure}
%%====================================
As we can observe in Fig. \ref{fig:sigmax}, the shape of the NLO cross section as a function of these three variables is rather similar for both the unpolarized  and polarized cases in the kinematical region, where the corresponding cross section is sizable. For the unpolarized result, $d\sigma/dX'_{\rm{Rec}}$ seems to be closer to $d\sigma/dX_{\rm{Real}}$ than $d\sigma/dX_{\rm{Rec}}$; in the polarized case, using the DSSV NLO parton distributions, the trend is the opposite. We also observe from Fig. \ref{fig:sigmax} that, with the kinematical cuts introduced above, the dominant range in $x$ for this observable is determined by $0.05\lesssim x \lesssim 0.15$.

%%==================================== 
\begin{figure}[htb]
\vspace{-0.3cm} 
\begin{center} 
\begin{tabular}{c} 
\epsfxsize=14truecm \epsffile{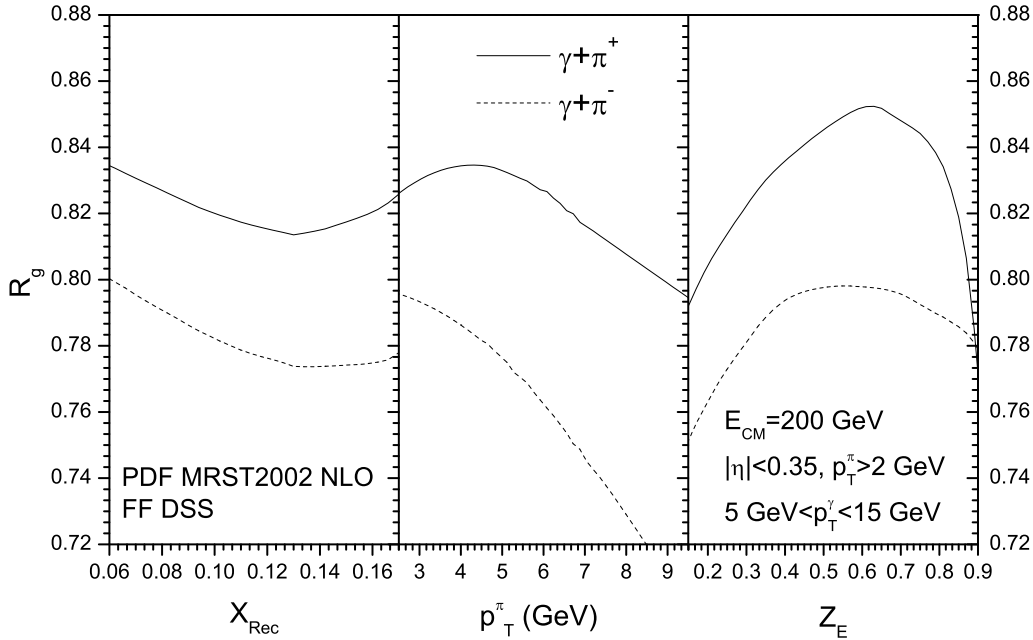}\\ 
\end{tabular} 
\end{center}
\vspace{-0.5cm} \caption{{\em \label{fig:rg} 
"`Gluonic ratio"' $R_g$ as a function of $X_{\rm{Rec}}$ (left), $p_{T}^{\pi}$ (center) and $Z_{E}$ (right), for unpolarized proton-proton collisions
at RHIC. We analyze separately the cross sections for positive (solid
lines) and negative (dashed lines) pions in the final state.}}
\end{figure}
%%====================================
In any case, comparing the shape of the cross section as a function of different variables is not enough to ensure that these variables have a strong correlation. Since the Monte Carlo implementation of the NLO calculation provides information about the kinematics of all the particles involved in the process, we can perform an event-by-event comparison of the values of $X_{\rm{Rec}}$, $X'_{\rm{Rec}}$ and $X_{\rm{Real}}$, weighting each possible configuration with its associated cross section. The result is presented in Fig. \ref{fig:corx}, where a gray-colored scale is used to indicate the probability of having an event with such values of the variables. A perfect correlation would be described by a black diagonal, as it occurs to LO accuracy: as we can observe $X_{\rm{Rec}}$ and $X'_{\rm{Rec}}$ are both strongly correlated with $X_{\rm{Real}}$. Using the definition of correlation for a nonuniform probability space, we find that the correlation coefficient
is larger than 0.76 for both pairs of variables $\left(X_{\rm{Real}},X_{\rm{Rec}}\right)$ and $\left(X_{\rm{Real}},X'_{\rm{Rec}}\right)$, when the bin size is set to $0.016$.

%%==================================== 
\begin{figure}[htb]
\vspace{-0.3cm} 
\begin{center} 
\begin{tabular}{c} 
\epsfxsize=14truecm \epsffile{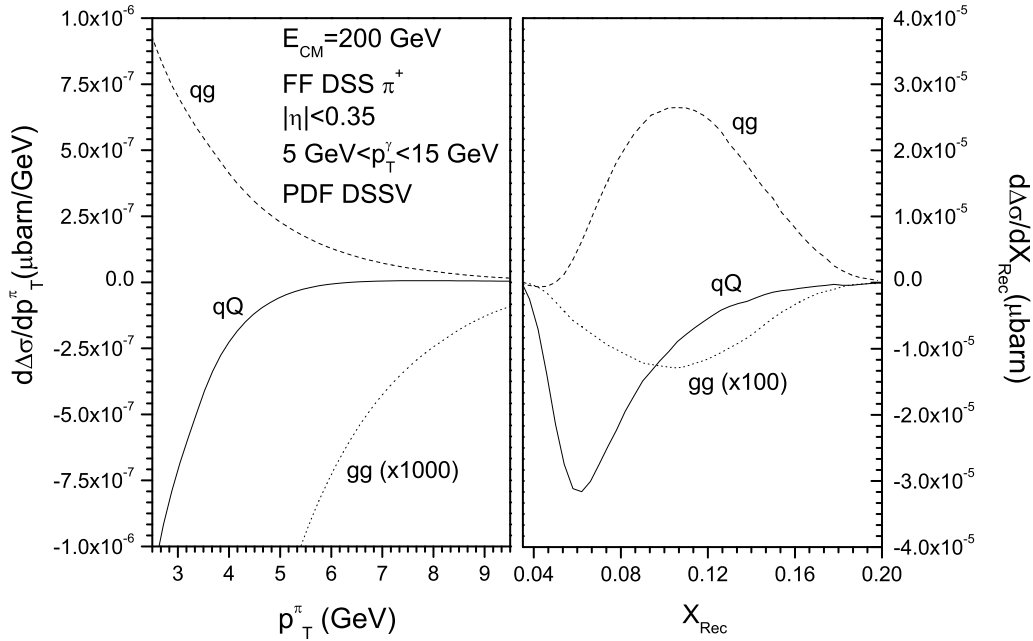}\\ 
\end{tabular} 
\end{center}
\vspace{-0.5cm} \caption{{\em \label{fig:rgpol} 
Contributions from the different initial state partonic channels to the polarized cross section as a function of  $p_{T}^{\pi}$ (left) and $X_{\rm Rec}$ (right), for polarized proton-proton collisions
at RHIC. Here $qQ$ corresponds to the sum of all channels initiated only by quarks or antiquarks.}}
\end{figure}
%%====================================
The reconstruction of the initial state momentum fractions might become very useful in order to extract information about parton distributions, and specially the polarized gluon density in polarized collisions. Regarding that, it is interesting to study the dependence of the cross section on the different partonic channels, to quantify the sensitivity on the gluon distribution. For that purpose we define the "`gluonic ratio"' as 
\beq R_g =
\frac{d\sigma_{gg}+d\sigma_{qg}}{d\sigma_{\rm{Total}}}  \, ,
\eeq 
which measures the relative importance of partonic processes initiated by at
least one gluon.

In Fig. \ref{fig:rg} we plot $R_g$ as a function of $p_{T}^{\pi}$, $Z_E$ and $X_{\rm{Rec}}$, where we distinguish between positive (solid lines) and negative (dashed lines) pions. Positively charged $u$ quarks and gluons dominate the inner structure of the proton, so it is more probable to have gluons or $u$ quarks in the final partonic state, which involves a higher probability of a fragmentation into positive pions and increases the sensitivity on the gluon distribution for $\pi^{+}+\gamma$ compared to  $\pi^{-}+\gamma$. As can be observed in Fig. \ref{fig:rg}, $R_g(\pi^{+})$
exceeds 0.80 for a wide kinematical range. Particularly, in the range of 3 GeV $\leq p^{\pi}_T \leq$ 10 GeV, more than 88 $\%$ of the events contain at least one gluon in the initial state,  confirming that hadron + photon production in hadronic collisions might provide a clear source of information on the gluon distribution. Of course, these ratios will considerably change in the polarized case due to the modification in the partonic cross sections as well as the different behavior and magnitude of the polarized distributions. As an example, we show in Fig. \ref{fig:rgpol} the contribution from each initial state partonic channel to the polarized cross section, for the (positively charged) pion transverse momentum and $X_{\rm Rec}$ distributions. It can be observed that the gluon contribution, mostly due to the $qg$ channel, is rather
sizable in the polarized case, even when a set with small gluon polarization like the DSSV set is used. In the next section we will analyze the sensitivity on $\Delta g$ in more detail.

%%==================================== 
\begin{figure}[htb]
\vspace{-0.3cm} 
\begin{center} 
\begin{tabular}{c} 
\epsfxsize=14truecm \epsffile{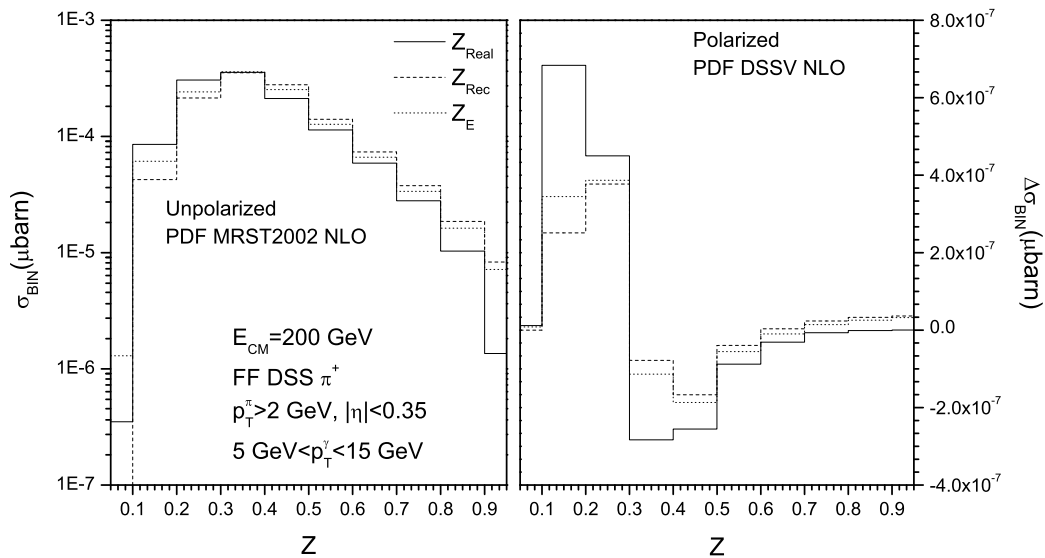}\\ 
\end{tabular} 
\end{center}
\vspace{-0.5cm} \caption{{\em \label{fig:sigz} 
NLO cross section as a function of $Z_{\rm{Real}}$ (solid),
$Z_{\rm{Rec}}$ (dash) and $Z_{E}$ (dot), for the unpolarized (left) and polarized (right) cases. Factorization and renormalization
scales are set to the default value.
}}
\end{figure}
%%====================================

On the other hand, in the case of heavy ion collisions it is of special interest to get a grip on the fragmentation variable, which would allow one to estimate the initial energy of the parton that undergoes hadronization and, eventually, loses energy when traveling through the high density matter formed in the collision.

In the previous section we already introduced the variable $Z_E$ (see Eq. (\ref{DefinicionZE})), which was employed in Ref. \cite{Adare:2009vd} to study partonic energy loss in Au-Au collisions at RHIC. At the Born level, $Z_E$ is equal to the true momentum fraction of the pion originated from the hadronization of parton 3, which we denote as $Z_{\rm{Real}}\equiv z $. Alternatively, one might define 
\beq Z_{\rm{Rec}} =
\frac{p^{\pi}_T}{p^{\gamma}_T} \, ,
\eeq
which is also equal to $Z_E$ at LO. As happened with $X_{\rm{Rec}}$ and $X'_{\rm{Rec}}$, $Z_{\rm{Rec}}$ and $Z_E$ can be used to approximate the real momentum fraction $Z_{\rm{Real}}$ in a NLO computation. To test the validity of this estimation, we plot in Fig. \ref{fig:sigz} the cross section as a function of $Z_{\rm{Real}}$ (solid), $Z_{\rm{Rec}}$ (dash) and $Z_E$ (dot), for the unpolarized (left) and polarized (right) cases. The three functions have a similar shape, although the discrepancies become larger for the polarized case in the region 0.15 $\leq z \leq$ 0.45, where the polarized cross section changes sign when the DSSV set of polarized PDFs is used. It is worth noticing
that the cross section is dominated by the fragmentation of a parton into a pion with a momentum fraction in the range $0.1\lesssim z \lesssim 0.6$.

%%==================================== 
\begin{figure}[htb]
\vspace{0.5cm} 
\begin{center} 
\begin{tabular}{c} 
\epsfxsize=14truecm \epsffile{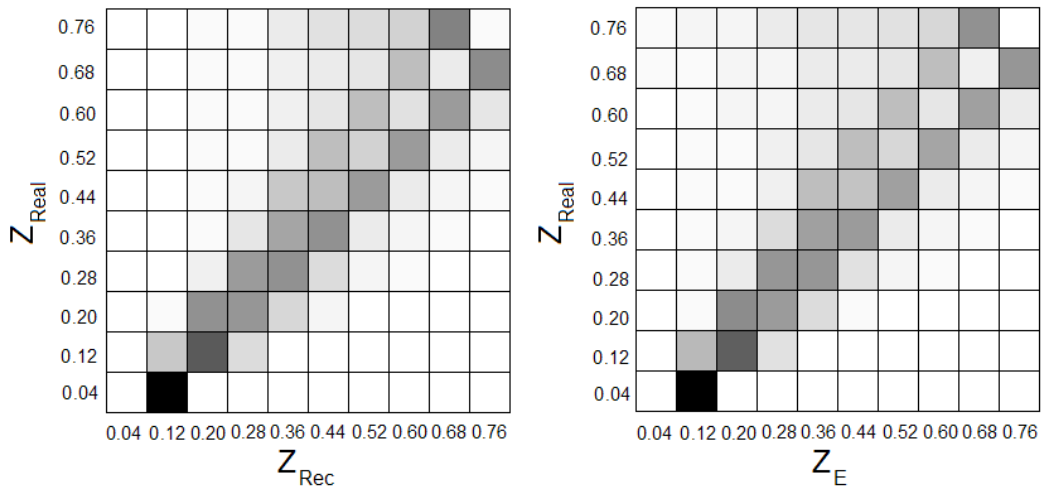}\\ 
\end{tabular} 
\end{center}
\vspace{-0.5cm} \caption{{\em \label{fig:corz} 
Correlation for $Z_{\rm{Rec}}$ (left) and $Z_{E}$ (right)
versus $Z_{\rm{Real}}$. As in Fig. \ref{fig:corx}, we use a gray scale picture to indicate
the probability of finding an event with a certain value of
$Z_{\rm{Real}}$ and $Z_{\rm{Rec}}$ (or $Z_{E}$): the darker the
bin, the larger the correlation between the variables.
}}
\end{figure}
%%====================================
As we did for the initial state momentum fractions, we also performed an event-by-event study to quantify the correlation of the pairs of variables $\left(Z_{\rm{Real}},Z_{\rm{Rec}}\right)$ and $\left(Z_{\rm{Real}},Z_E\right)$. Applying the same algorithm used for initial state momentum fraction, the correlation coefficient is 0.77 for $\left(Z_{\rm{Real}},Z_{\rm{Rec}}\right)$ and 0.71 for $\left(Z_{\rm{Real}},Z_E\right)$, when the bin size is set to $0.08$. Therefore, we conclude that $Z_{\rm{Rec}}$ allows for a more accurate reconstruction of $Z_{\rm{Real}}$ at NLO accuracy. In spite of having a high correlation coefficient, the reconstruction is not exact and both approximations show a bifurcation in their trends as $z$ increases, due to the fall of the cross section. This situation can be observed in Fig. \ref{fig:corz}, where we can appreciate that the deviations from the exact values are anyway typically smaller than 15$\%$. These correlations can be improved by setting stronger cuts on the transverse momentum of the photon, like $p_T^{\gamma}>10$ GeV, but with the corresponding loss in cross section.
This level of reconstruction might help to obtain a better understanding of
the energy loss of partons when traveling through the dense medium created in heavy ion collisions and, particularly, to test whether the parton energy loss can be accounted for via an effective modification of the fragmentation functions. We have analyzed other possible definitions for the reconstructed version of the variables $x$ and $z$, all of them equivalent to LO accuracy. We do not find sensible differences in the corresponding correlations with the real variables compared to those already presented here.

%%==================================== 
\begin{figure}[htb]
\vspace{-0.3cm} 
\begin{center} 
\begin{tabular}{c} 
\epsfxsize=14truecm \epsffile{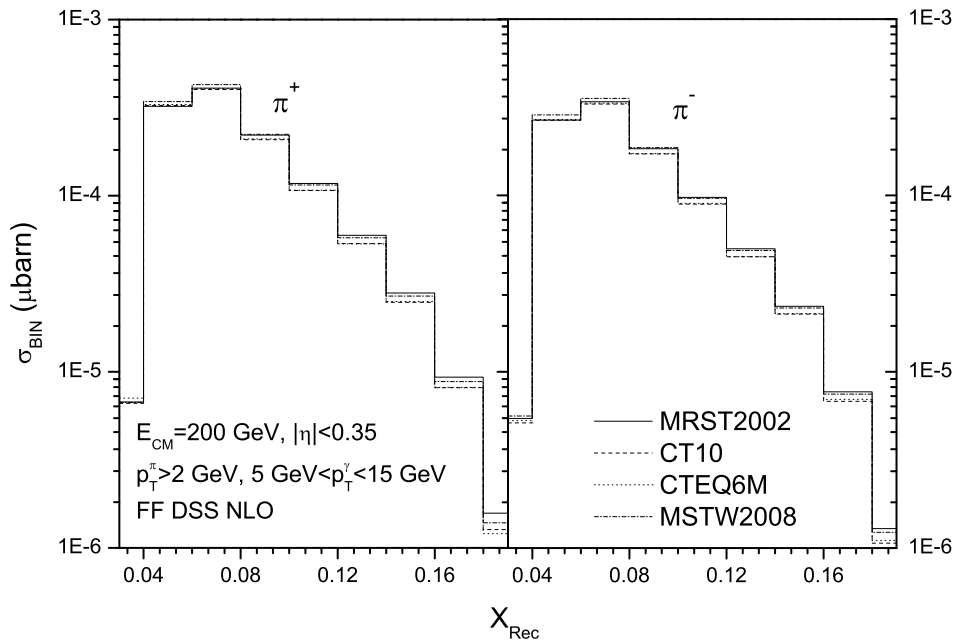}\\ 
\end{tabular} 
\end{center}
\vspace{-0.5cm} \caption{{\em \label{fig:updf} 
$X_{\rm Rec}$ distribution for a photon accompanied by a positively (left-side plot) and negatively (right-side plot) charged pion computed using different set of parton distributions: MRST2002 (solid lines), CTEQ6M (dotted lines), CT10 (dashed lines) and MSTW2008 (dotted-dashed lines). }}
\end{figure}
%%====================================

%%==================================== 
\begin{figure}[htb]
\vspace{-0.3cm} 
\begin{center} 
\begin{tabular}{c} 
\epsfxsize=14truecm \epsffile{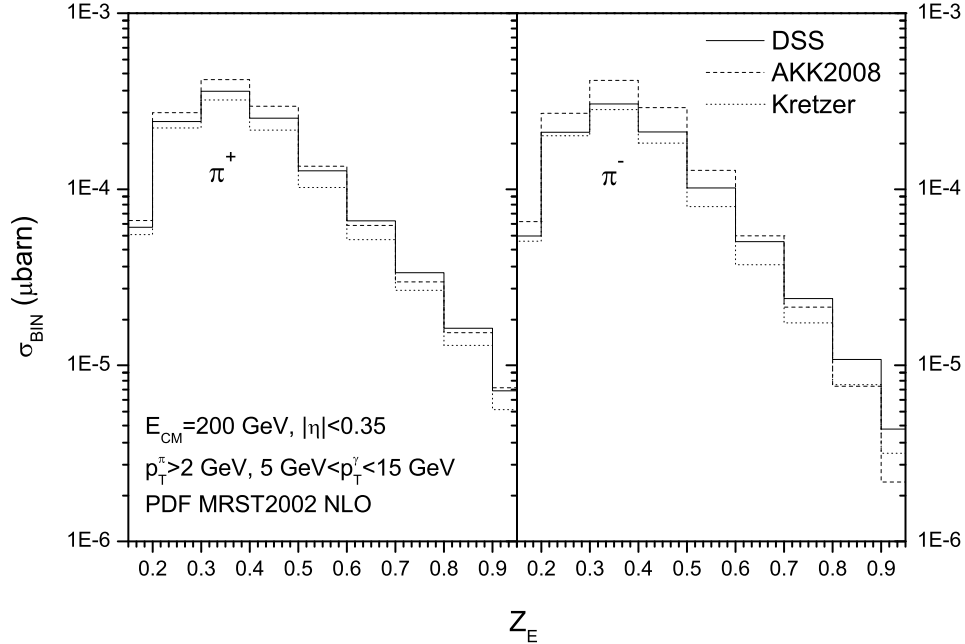}\\ 
\end{tabular} 
\end{center}
\vspace{-0.5cm} \caption{{\em \label{fig:ff} 
$Z_{E}$ distribution for a photon accompanied by a positively (left-side plot) and negatively (right-side plot) charged pion computed using different set of fragmentation functions: DSS (solid lines), AKK (dashed lines) and Kretzer (dotted lines).}}
\end{figure}
%%====================================

Before moving to the phenomenological analysis of the asymmetries, we would like to quantify the possible uncertainties on the analyzed observables due to the unpolarized parton distributions and fragmentation functions. As stated before, we rely on MRST2002 as the default set because the DSSV analysis is based on those distributions. In Fig. \ref{fig:updf} we show the $X_{\rm Rec}$ distribution for a photon accompanied by a positively (left-side plot) and negatively (right-side plot) charged pion computed using different sets of parton distributions: MRST2002, CTEQ6M \cite{cteq6}, CT10 \cite{ct10} and MSTW2008 \cite{mstw2008}. While the differences become more visible in the kinematical range where the cross section becomes rather small, in general we do not observe variations large enough to spoil the interpretation of the measured asymmetries. This is not unexpected, since modern sets of PDFs provide distributions that agree within a few percent between them. The situation is more complicated when one turns to the fragmentation functions, since those distributions still suffer from much larger uncertainties. In order to study the dependence on the set of fragmentation functions used, we plot in Fig. \ref{fig:ff} the $Z_E$ distribution for both positively and negatively charged pions obtained using the DSS \cite{deFlorian:2007hc}, AKK \cite{akk} and Kretzer \cite{Kretzer} sets of fragmentation functions. As anticipated, the variations observed here are more sizable than the ones found for the parton distributions. Nevertheless, we will demonstrate in the next section that the dependence on the fragmentation functions, which contribute in a similar way to both polarized and unpolarized cross sections, partially cancels when computing the corresponding asymmetries.

%%==================================== 
\begin{figure}[htb]
\vspace{-0.3cm} 
\begin{center} 
\begin{tabular}{c} 
\epsfxsize=13truecm 
\epsffile{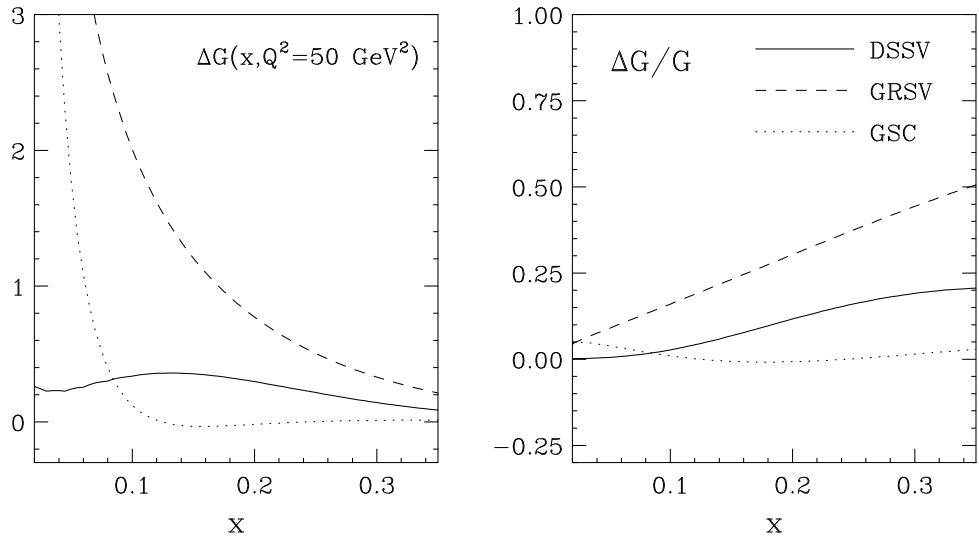}\\ 
\end{tabular} 
\end{center}
\vspace{-0.5cm} \caption{{\em \label{fig:pdf} Polarized gluon density at $Q^2=50 \, \rm{GeV}^2$ from different sets of polarized PDFs 
(left) and their ratios to the unpolarized distribution (right). 
}}
\end{figure}
%%====================================

\section{Spin asymmetries}
As a way to directly analyze the sensitivity of the process on the polarized
gluon distribution, we will compute the NLO asymmetries with 3 different sets of spin-dependent densities: DSSV \cite{deFlorian:2008mr}, GRSV (standard) \cite{grsv} and GS-C \cite{gs}. The corresponding NLO distributions at $Q^2=50$ GeV$^2$, a typical scale for this process, are shown in the left-side of Fig. \ref{fig:pdf}. As can be observed, the expectations from the three sets are quite different. Here, the DSSV distribution corresponds to the best fit from the latest global analysis of all polarized data \cite{deFlorian:2008mr}, while the GRSV set can be considered as an (overestimate of the) "`upper bound"' for the allowed range of gluon densities, even though this set has already been ruled out by the available data. The GS-C set provides a distribution compatible with the requirement of a small gluon polarization in the range $0.05\lesssim x\lesssim 0.3$ but with a node in that region and a very different behavior at smaller $x$ compared to the DSSV set. The right side in Fig. \ref{fig:pdf} shows the ratio between the corresponding polarized distribution and the unpolarized MRST2002 set. In the quark sector, the dominant distributions are very similar among the different sets. Therefore, we can expect that any differences between predictions for the asymmetries that are
found when using different polarized parton density sets are to be attributed to the sensitivity of the observable to $\Delta g$.
%%==================================== 
\begin{figure}[h]
\vspace{-0.3cm} 
\begin{center} 
\begin{tabular}{c} 
\epsfxsize=14truecm \epsffile{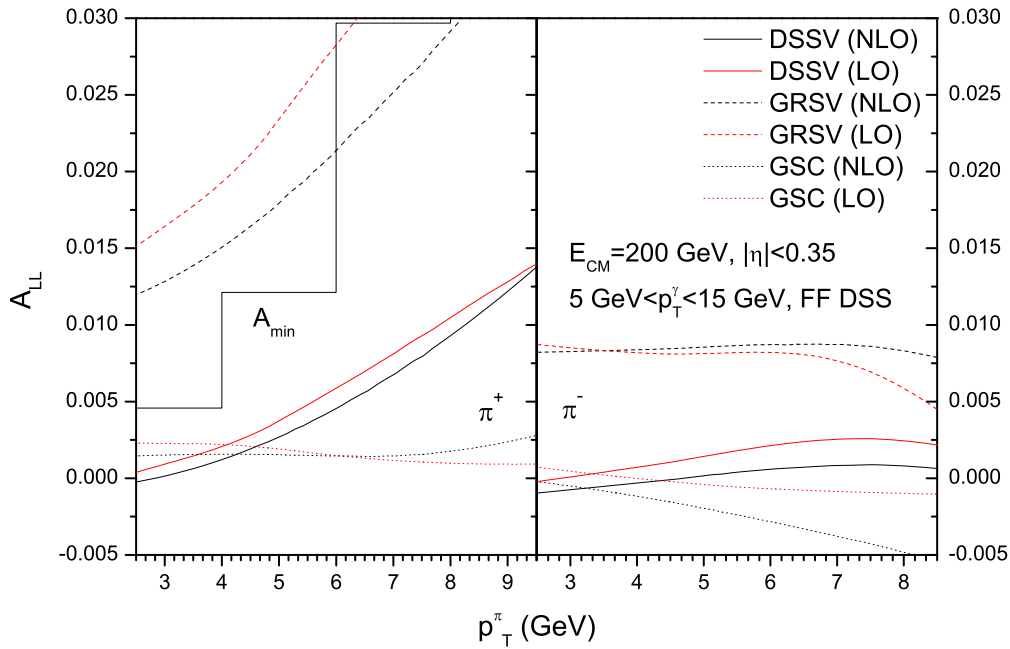}\\ 
\end{tabular} 
\end{center}
\vspace{-0.5cm} \caption{{\em \label{fig:Apt} 
Expected NLO and LO asymmetries for $\pi^+$ (left) and $\pi^-$ (right) production at RHIC in terms of the transverse momentum of the pion for different sets of polarized PDFs: DSSV, GS-C and GRSV. The histogram in the left-side plot represents the value $A_{min}$ as defined in Eq. (\ref{stat}). }}
\end{figure}
%%====================================

For the presentation of our phenomenological results we will concentrate on the
production of positively charged pions, since that observable provides, in the polarized case, a stronger sensitivity on the polarized gluon distribution. We  
start with the expected asymmetries by looking first at  $\pi^{\pm}+\gamma$ production in terms of the transverse momentum of the pion as shown in Fig. \ref{fig:Apt}. As expected, the asymmetries for the DSSV and GS-C distributions turn out to be rather small. When a set with a larger gluon distribution  is considered, like GRSV, the asymmetries increase to the $\sim 2\%$ level for $\pi^+$ production. A rough estimate of the minimum value of the asymmetry observable at RHIC due to statistics can be obtained by the well-known formula 
\beqn 
A_{min} = \frac{1}{P^2} \frac{1}{\sqrt{2 \sigma {\cal L} \epsilon}} \, ,
\label{stat} 
\eeqn
where ${\cal L}$ is the integrated luminosity, $P$ is the polarization of the beam, and the factor $\epsilon \le 1$ accounts for experimental efficiencies; $\sigma$ is the unpolarized cross section integrated over a small range in transverse momentum of the pion. The quantity defined in Eq. (\ref{stat}) is plotted (histogram) in the left side of Fig. \ref{fig:Apt}, for $\epsilon=1$, $P = 0.7$, ${\cal L} = 100$ pb$^{-1}$, and a $p_T$-bin size of 2 GeV. From the value obtained with these particular set of parameters one can notice that the minimally observable asymmetry at RHIC is of the order of magnitude of the expected value, and typically larger than the one obtained using the DSSV set. A similar situation has been observed for the asymmetries already measured at RHIC in the case of jet or single-inclusive hadron production, which turned out to be the key measurements to determine the polarized gluon distribution so far \cite{deFlorian:2008mr}. Therefore, one might expect that the measurement of $\pi^+ + \gamma$ at RHIC will also play an important role in future global analysis.
As anticipated in the unpolarized analysis (see Fig. \ref{fig:rgpol}), the sensitivity on the gluon distribution drops considerably for $\pi^-$, an effect which is reflected in the decrease of the corresponding asymmetry. Nevertheless, the separate measurement of asymmetries for positively and negatively charged pions can provide extra information on the polarized quark flavor structure once incorporated in a global analysis of all observables. It is also worth noticing from Fig. \ref{fig:Apt} that the NLO corrections in general tend to reduce the asymmetries, but the effect is nontrivial and strongly depends on the transverse momentum of the pion and the polarized set of parton distributions used.

%%==================================== 
\begin{figure}[htb]
\vspace{-0.3cm} 
\begin{center} 
\begin{tabular}{c} 
\epsfxsize=14truecm \epsffile{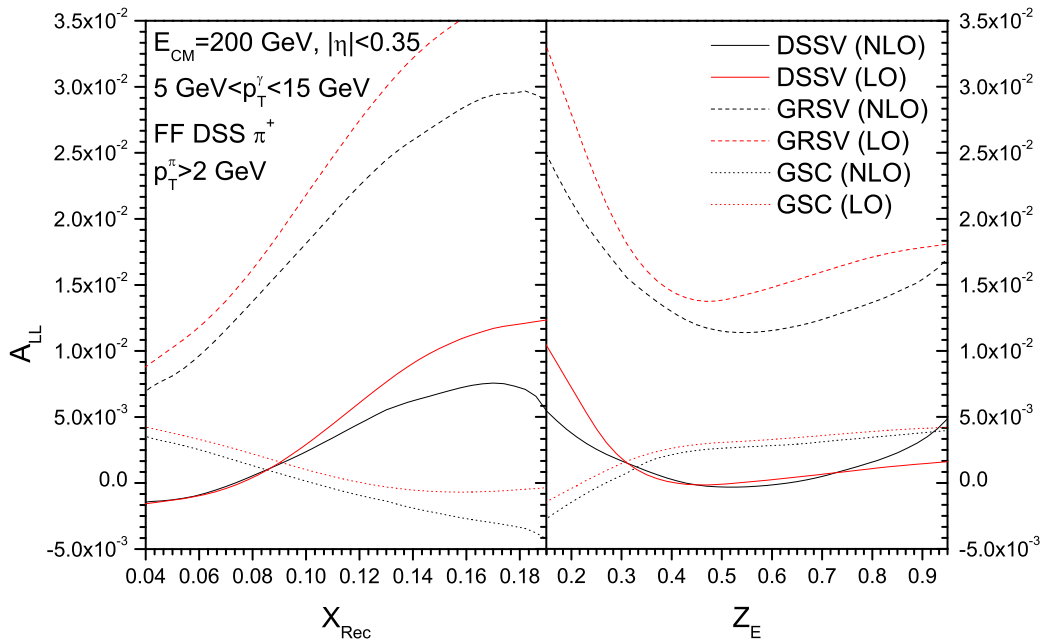}\\ 
\end{tabular} 
\end{center}
\vspace{-0.5cm} \caption{{\em \label{fig:Ax} 
Same as Fig. \ref{fig:Apt} but in terms of the scaling variable $X_{\rm{Rec}}$ (left) and $Z_E$ (right).
}}
\end{figure}
%%====================================

A similar study is presented in Fig. \ref{fig:Ax}, where we analyze the corresponding asymmetries for $\pi^+ +\gamma$ production in terms of the scaling variables $X_{\rm{Rec}}$ (left) and $Z_E$ (right). In particular, the left-side plot in Fig. \ref{fig:Ax} reflects the $x$-shape and order of the
curves for the $\Delta G/G$ ratio plotted in the right side of Fig. \ref{fig:pdf}, confirming the strong sensitivity of this observable on the gluon polarization. 

%%==================================== 
\begin{figure}[htb]
\vspace{-0.3cm} 
\begin{center} 
\begin{tabular}{c} 
\epsfxsize=14truecm \epsffile{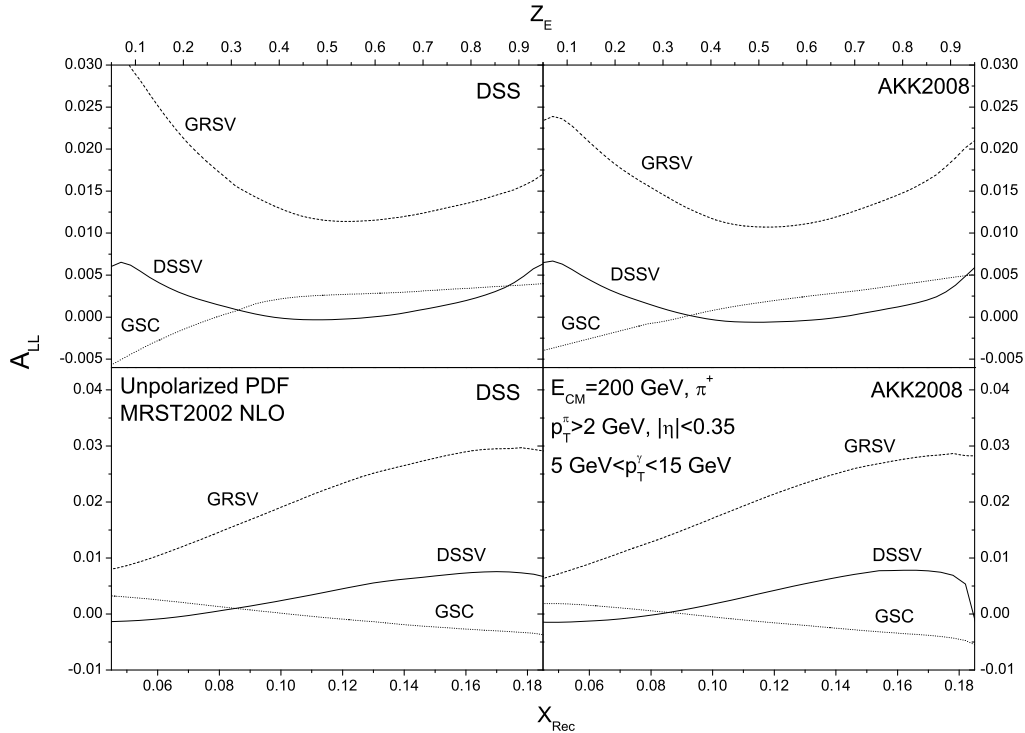}\\ 
\end{tabular} 
\end{center}
\vspace{-0.5cm} \caption{{\em \label{fig:Aff} 
NLO asymmetries in terms of the scaling variable $Z_E$ (up) and $X_{\rm{Rec}}$ (down). The left-side plots correspond to results obtained using the default DSS set of fragmentation functions while those in the right were obtained using the AKK2008 set.}}
\end{figure}
%%====================================
As discussed in the previous section, one issue that can veil the possible extraction of information about the polarized parton distributions arises from the uncertainty in the fragmentation functions. A comparison at the level of the asymmetries in terms of both scaling variables $X_{\rm{Rec}}$ and $Z_E$ is presented in Fig. \ref{fig:Aff}, where we confront the results obtained using the DSS (left-side plots) and AKK2008 (right-side plots) of fragmentation functions to NLO accuracy. For most of the kinematical range that can be covered by the RHIC experiment, the dependence in the fragmentation functions, quite visible at the level of the cross section in Fig. \ref{fig:ff}, almost cancels in the asymmetry ratios, with the exception of the end points in $Z_E$, where fragmentation functions are hardly constrained by the available data.

%%==================================== 
\begin{figure}[htb]
\vspace{-0.3cm} 
\begin{center} 
\begin{tabular}{c} 
\epsfxsize=14truecm \epsffile{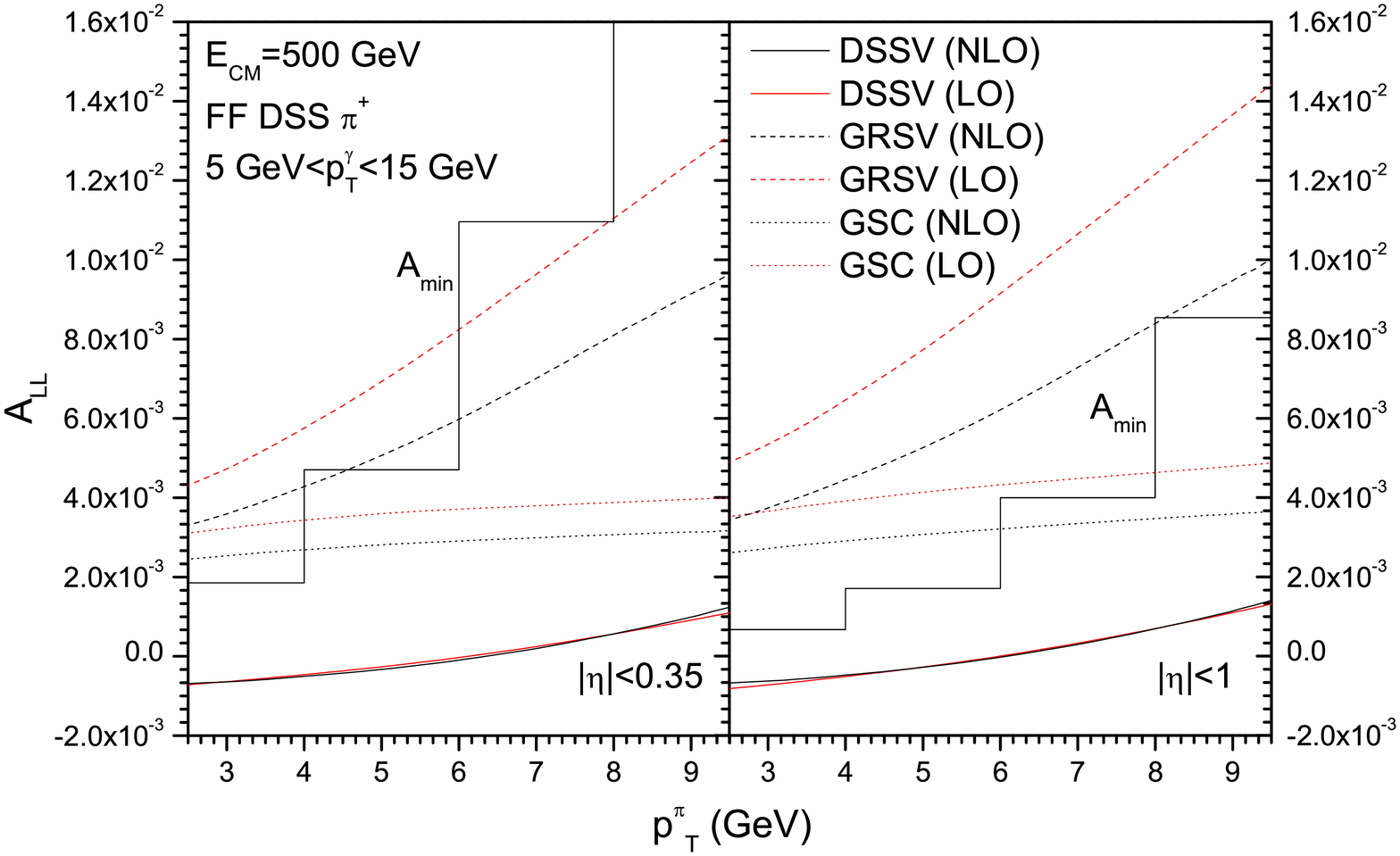}\\ 
\end{tabular} 
\end{center}
\vspace{-0.5cm} \caption{{\em \label{fig:pt500} 
Expected asymmetries for 
$\pi^+$ production at RHIC in terms of the transverse momentum of the pion
for center-of-mass energy $\sqrt{s}=500 \, {\rm GeV}$ and PHENIX (left side) and STAR (right side) kinematics. The histogram represents the value $A_{min}$ as defined in Eq. (\ref{stat}).}}
\end{figure}
%%====================================

The results presented so far correspond to a center-of-mass energy $\sqrt{s}=200$ GeV and were restricted to the kinematics accesible to the PHENIX experiment. In the following we discuss the extension of the phenomenological analysis of the asymmetries to a center-of-mass energy $\sqrt{s}=500$ GeV and also account for possible measurements at STAR, which extends the rapidity coverage to $|\eta |\le 1$. 

In Fig. \ref{fig:pt500} we present the asymmetry for $\pi^+$ production at $\sqrt{s}=500$ GeV in terms of the transverse momentum of the pion for PHENIX (leftside plot) and STAR (rightside plot). As can be observed, the asymmetries show the same pattern found for $\sqrt{s}=200$ GeV but almost a factor of 2 smaller in magnitude. For the same value of transverse momentum, the cross section becomes sensitive to smaller values of $x$ when the center-of-mass energy increases. Since the  polarized parton distributions fall faster than the unpolarized ones at small $x$, that results in the reduction observed in the asymmetries. Also, at smaller $x$ the quark-initiated channels play a more important role in the polarized cross section and tend to reduce the gluon sensitivity of the asymmetry.  

%%==================================== 
\begin{figure}[htb]
\vspace{-0.3cm} 
\begin{center} 
\begin{tabular}{c} 
\epsfxsize=14truecm \epsffile{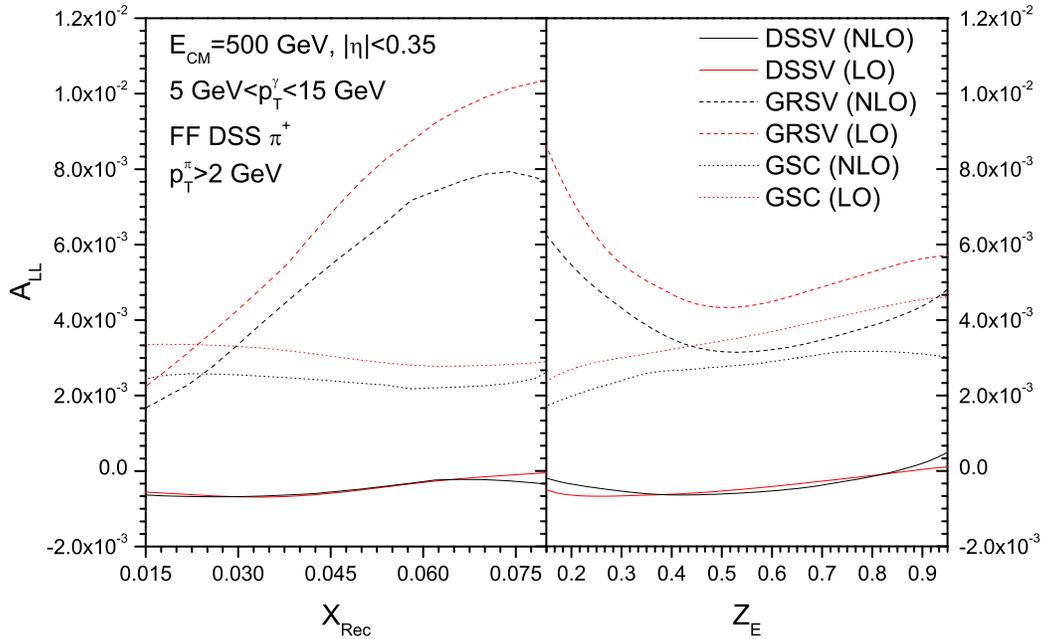}\\ 
\end{tabular} 
\end{center}
\vspace{-0.5cm} \caption{{\em \label{fig:p500} 
Same as Fig. \ref{fig:Ax} for center-of-mass energy $\sqrt{s}=500 \, \rm{GeV}$ and PHENIX kinematics.}}
\end{figure}
%%====================================

For this particular observable one can not appreciate large deviation between the predictions for PHENIX and STAR. The integration over a larger range in rapidity at STAR provides a greater cross section, with the corresponding increase in statistics,  but that normalization effect almost cancels in the asymmetry. We also include in Fig. \ref{fig:pt500} the estimate of the minimum value of the asymmetry observable at RHIC for both experiments obtained using Eq. (\ref{stat}), assuming an integrated luminosity of ${\cal L} = 300$ pb$^{-1}$. We observe that, despite the decrease in the asymmetries because of the larger center-of-mass energy, the analyzing power of the observable actually improves due to the much higher statistics that can be collected at this energy for both experiments.

%%==================================== 
\begin{figure}[htb]
\vspace{-0.3cm} 
\begin{center} 
\begin{tabular}{c} 
\epsfxsize=14truecm \epsffile{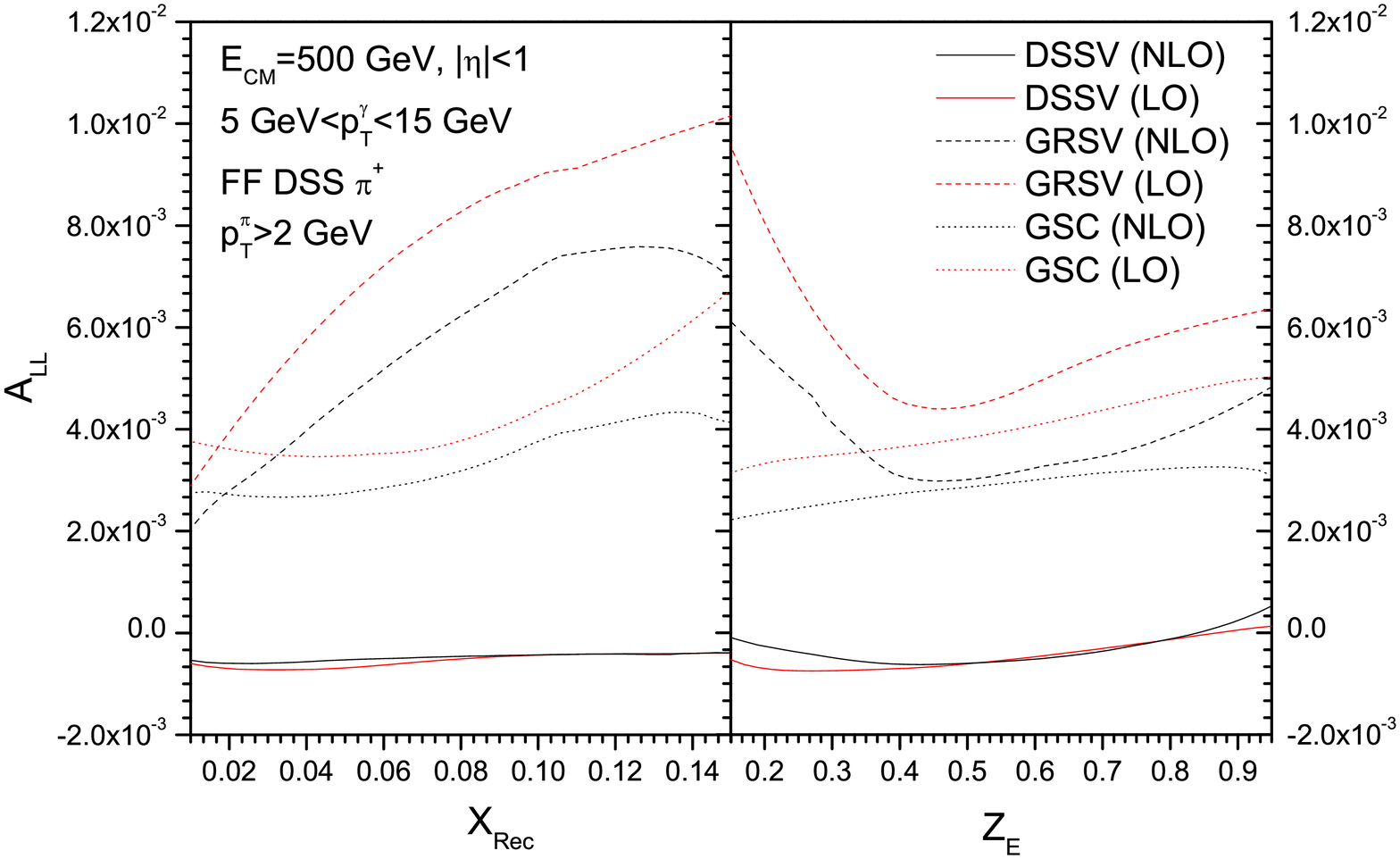}\\ 
\end{tabular} 
\end{center}
\vspace{-0.5cm} \caption{{\em \label{fig:s500} 
 Same as Fig. \ref{fig:p500} for STAR kinematics.}}
\end{figure}
%%====================================

Finally, we present in Figs. \ref{fig:p500} and \ref{fig:s500} the asymmetries for $\pi^+$ production in terms of the scaling variables $X_{\rm Rec}$ (left-side plots) and $Z_{E}$ (right-side plots) for PHENIX and STAR kinematics, respectively. In the particular case of the $X_{\rm Rec}$ distribution, we observe that one can span values of $x$ about a factor of 2 smaller than at $\sqrt{s}=200$ GeV. Furthermore, one can appreciate subtle differences between the results for PHENIX and STAR, mostly due to the fact that the larger rapidity coverage of STAR slightly broadens the range in the momentum fraction of the initial state partons that contribute to the observable.

\section{Conclusions}

In this paper we presented the NLO corrections for the unpolarized and polarized cross sections for the production of a hadron and a back-to-back isolated photon. 

The corrections are found to be nontrivial: `$K$-factors' are larger for the unpolarized cross section than for the polarized one, typically resulting in a reduction of the asymmetry at NLO. It is shown that the perturbative stability of the hadron plus photon cross section improves after including the NLO contributions, particularly when analyzing the dependence on the factorization 
scales, but the total scale dependence remains rather sizable due to
the implementation of isolation cuts for the photon.

We confirmed that "`naive"' LO relations between the photon and pion momentum
retain a strong correlation to the true partonic fractions which are the arguments of the parton distributions ($x_1$ and $x_2$) and fragmentation functions ($z$) when NLO effects are taken into account. Counting with more precise distributions on the fragmentation fraction $z$ might permit one to obtain a better understanding of the energy loss of partons when traveling through the dense medium created in heavy ion collisions. 

Finally, the possibility of looking at a charged pion accompanied by a back-to-back isolated photon is studied phenomenologically in detail for polarized collisions, as feasible at RHIC. We find that the asymmetries for $\pi^+$ production, in terms of both dimensionless variables $X_{\rm Rec}$ and $Z_E$, are sensitive to the polarized gluon density in the range $0.05\lesssim x\lesssim 0.15$ and might contribute to a better understanding of the spin content of the proton.

%%%%%%%%%%%%%%%%%%%%%%%%%%%%%%%%%%%%%%%%%%%%%%%%%%%%%%%%%%%%%%%%%%%%%%%  
  \subsection*{Acknowledgments}

This work was partially supported by UBACYT, CONICET, ANPCyT and the EU Grant PITN-GA-2010-264564.
%%%%%%%%%%%%%%%%%%%%%%%%%%%%%%%%%%%%%%%%%%%%%%%%%%%%%%%%%%%%%%%%%%%%%%%  

%\begin{center}
%\large{ \bf References}
%\end{center}

\end{document}